\documentclass[aps,showpacs,superscriptaddress,twocolumn]{revtex4-1}
\usepackage{mathrsfs}
\usepackage{amsfonts}
\usepackage{amssymb}
\usepackage{amsmath}
\usepackage{graphicx}
\usepackage{dsfont}
\usepackage[usenames,dvipsnames]{color}
\usepackage[colorlinks=true,citecolor=blue,linkcolor=magenta]{hyperref}
\usepackage{ulem}

\newcommand{\abs}[1]{| #1 |}

\newcommand{\ket}[1]{\vert{ #1 }\rangle}

\newcommand{\ketbra}[2]{\vert #1 \rangle \langle #2 \vert}

\begin{document}

\title{Bidirectional and passive optical field to microwave field quantum converter with high bandwidth}

\author{Mingxia Huo}
\address{Beijing Computational Science Research Center, Beijing 100193, China}

\begin{abstract}
The conversion between microwave photons and optical photons with quantum coherence is important for quantum communication and computation. In this paper, we report a proposal using an ensemble of atoms coupled to microwave and optical resonators. Input photons to one resonator are converted into output photons in the other resonator without active operation. Usually the conversion is only optimized at certain frequency. In our proposal, we find that the efficiency is almost a constant and can be close to $100\%$ in a large interval of frequency, i.e.~a high-bandwidth conversion can be realized with our proposal.
\end{abstract}

\maketitle

\section{Introduction}

Quantum state transfer from one place to another is crucial for quantum communication. Photons are natural flying qubit carriers. The telecommunication optical fibre makes photons capable to distribute over a long distance. For quantum information storage and processing, quantum states encoded into superconducting circuits or electron spins that operate at microwave wavelengths are usually considered. One future quantum network will consist of optical fibres and long-term memories. It will require interfaces that enable transfers between photons in different wavelengths. During the conversion, the quantum coherence is preserved. 

Quantum conversion between microwave and optical frequencies remains a challenge. The development of such a converter requires highly efficient and reversible mapping of microwave to optical photons at the quantum level. Several approaches of the converter have been explored. The converter using mechanical resonators have been proposed in theory~\cite{naeini:11,regal:11}, and demonstrated in experiments~\cite{bochmann:13,bagci:14,andrews:14}. In these systems, both optical and microwave resonators are parametrically coupled to a mechanical oscillator. This system is appealing for quantum frequency conversion. The reason is that the mechanical oscillator can couple to light of any frequency. Therefore, through varying the resonant frequencies of resonators, the frequency of the input signal can be modulated into a specific range of the spectrum. The major obstacles are the thermal motion of the mechanical oscillator, and quantum back-action noise of fields in each electromagnetic resonator~\cite{hill:12}. To have coherent, lossless, and noiseless conversion, the rather low frequency mechanical resonator needs to be cooled to its quantum ground state. The other approach is utilizing spin ensembles as a medium to couple microwave and optical photons~\cite{marcos:10,barzanjeh:12,hafezi:12,maxwell:13,brien:14,williamson:14,gonzalvo:15,gonzalvo:17,welinski:18}. Among the proposals based on spin ensembles, rare-earth doped crystals are very attractive. This is due to their high spin tuning rate and long optical and spin coherence time~\cite{probst:13}. A promising choice is rare-earth element erbium with telecom-wavelength and microwave transitions~\cite{brien:14,gonzalvo:15}. However, strong coupling between a microwave resonator and erbium spins has not been demonstrated. The challenge also lies in the fast dephasing rates and the large inhomogeneous broadening of the spin ensembles~\cite{kurucz:11,diniz:11}.

A recent microwave-to-optical conversion proposal suggests using the erbium atoms doped crystal~\cite{williamson:14}. It acts as an ensemble of $\Delta$-type three-level atoms~\cite{zhou:13} to couple with a microwave resonator, an optical resonator, and a coherent driving field~\cite{williamson:14}. A shielded loop-gap resonator and a Fabry-Perot resonator are considered as the microwave and optical resonators, respectively. The doped erbium in yttrium orthosilicate sits in the middle hole of the loop-gap resonator. Here the magnetic field is concentrated in. In such a proposal, the bandwidth is limited by the effective interaction between the microwave resonator mode and the optical resonator mode. This is a result of the third-order perturbation.

In this paper, we adapt the proposal in Ref.~\cite{williamson:14}. We consider the same approach that an ensemble of atoms is coupled to microwave and optical resonators. However, two modes are not directly coupled via the effective interaction. The conversion is realized by coupling two modes to an effective mode of atomic excitations. We focus on the case that three modes are on resonance. We find that in our proposal the conversion efficiency can reach $100\%$ at certain frequencies. The input microwave photons can be completely converted into optical photons while preserving quantum coherence, and vice versa. The conversion does not require any active control at the arrival of the input field. Usually, the conversion efficiency decreases when the frequency of the input field moves away from the optimal frequency. The bandwidth is the frequency interval, in which the conversion efficiency is sufficiently high. In our proposal, we find that the efficiency is almost a constant in a big interval of the frequency. The typical width of the high-efficiency interval is the effective coupling between the microwave resonator and the ensemble of atoms. It is proportional to square root of the number of atoms. Within the interval, the conversion efficiency can reach $99.9\%$ for any frequency of the input field. Therefore, high bandwidth conversion can be realized using our proposal.

\section{The system and Hamiltonian}

\begin{figure}[tbp]
\centering
\includegraphics[width=1\linewidth]{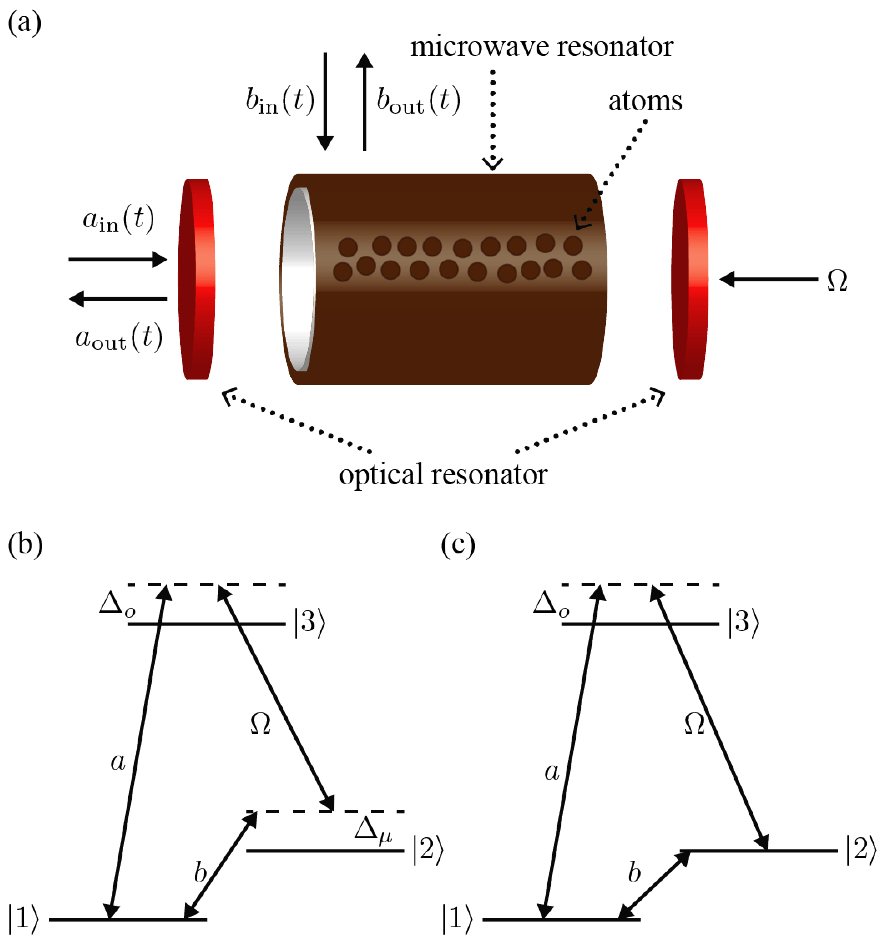}
\caption{(Colour online) (a) The setup under consideration. Atoms are coupled to an optical resonator and a microwave resonator. (b) The scheme with large detunings $\abs{\Delta_{o}} \gg \sqrt{N} \abs{g_{o}}, \abs{\Omega}$ and $\abs{\Delta_{\mu}} \gg \sqrt{N} \abs{g_{\mu}}, \abs{\Omega}$. Here, $N$ is the number of atoms. (c) The setup with a resonant transition, i.e.~$\abs{\Delta_{o}} \gg \sqrt{N} \abs{g_{o}}, \abs{\Omega}$ but $\Delta_{\mu}$ vanishes.}
\label{fig:setup}
\end{figure}

The optical field to microwave field converter considered in this paper is an ensemble of atoms coupled to an optical resonator and a microwave resonator. The conversion between the two fields is realized by transitions between three atomic levels (see Fig.~\ref{fig:setup}). The optical field couples levels $\ket{1}$ and $\ket{3}$. The microwave field couples levels $\ket{1}$ and $\ket{2}$. A coherent field drives the transition between levels $\ket{2}$ and $\ket{3}$. In order to convert an optical photon into a microwave photon, the optical photon is absorbed in the transition from $\ket{1}$ to $\ket{3}$. Then an optical photon into the coherent field and a microwave photon into the microwave resonator are emitted intermediated by $\ket{2}$. And the atomic state goes back to $\ket{1}$. The same system can also convert microwave photons into optical photons following the reverse route. Such a level structure can also be used for quantum routing of single photons~\cite{zhou:13}.

As shown in Fig.~\ref{fig:setup}, in the interaction picture the converter is described by the Hamiltonian
\begin{eqnarray}
H &=& \sum_{k} \Delta_{o,k} \sigma_{33,k} + \sum_{k} \Delta_{\mu,k} \sigma_{22,k} \notag \\
&&+ \sum_{k} (g_{o,k} a^{\dagger}\sigma_{13,k} + h.c.) + \sum_{k} (g_{\mu,k} b^{\dagger} \sigma_{12,k} + h.c.) \notag \\
&&+ \sum_{k} \Omega_{k} (\sigma_{23,k} + h.c.).
\end{eqnarray}
Here, $\sigma_{ij,k} = \ketbra{i}{j}_k$ is an operator of the $k$-th atom, $a$ is the annihilation operator of photons in the optical resonator, and $b$ is the annihilation operator of photons in the microwave resonator. The transition $\ket{1} \leftrightarrow \ket{3}$ is coupled to the optical resonator with the strength $g_{o,k}$ and detuning $\Delta_{o,k}$, and the transition $\ket{1} \leftrightarrow \ket{2}$ is coupled to the microwave resonator with the strength $g_{\mu,k}$ and detuning $\Delta_{\mu,k}$. We choose the frequency of the coherent field at $\omega_c = \omega_o - \omega_\mu$, where $\omega_o$ is the frequency of the optical resonator, and $\omega_\mu$ is the frequency of the microwave resonator. $\Omega_{k}$ is the Rabi frequency of the transition $\vert 2\rangle \leftrightarrow \vert 3\rangle$ driven by the coherent field.

The setup that the transition between $\ket{2}$ and $\ket{3}$ is detuned [see Fig.~\ref{fig:setup}(b)] is proposed in Ref.~\cite{williamson:14}. In this paper, we consider the setup that the transition between $\ket{2}$ and $\ket{3}$ is resonating with the microwave resonator [see Fig.~\ref{fig:setup}(c)], and we show that resonant setup results in a much larger bandwidth.

In the detuned setup, atomic excitations to $\ket{2}$ and $\ket{3}$ can be adiabatically eliminated, then effectively optical and microwave fields are directly coupled. The effective Hamiltonian is $H_{a,b} = Sa^\dag b + h.c.$. Here, the effective coupling strength is $S = \sum_k \frac{\Omega_k g_{\mu,k} g_{o,k}^*}{\Delta_{\mu,k} \Delta_{o,k}}$, which is a result of third-order perturbabtion and limits the bandwidth of the converter~\cite{williamson:14}.

\section{Resonant setup and effective Hamiltonian}

\begin{figure}[tbp]
\centering
\includegraphics[width=1\linewidth]{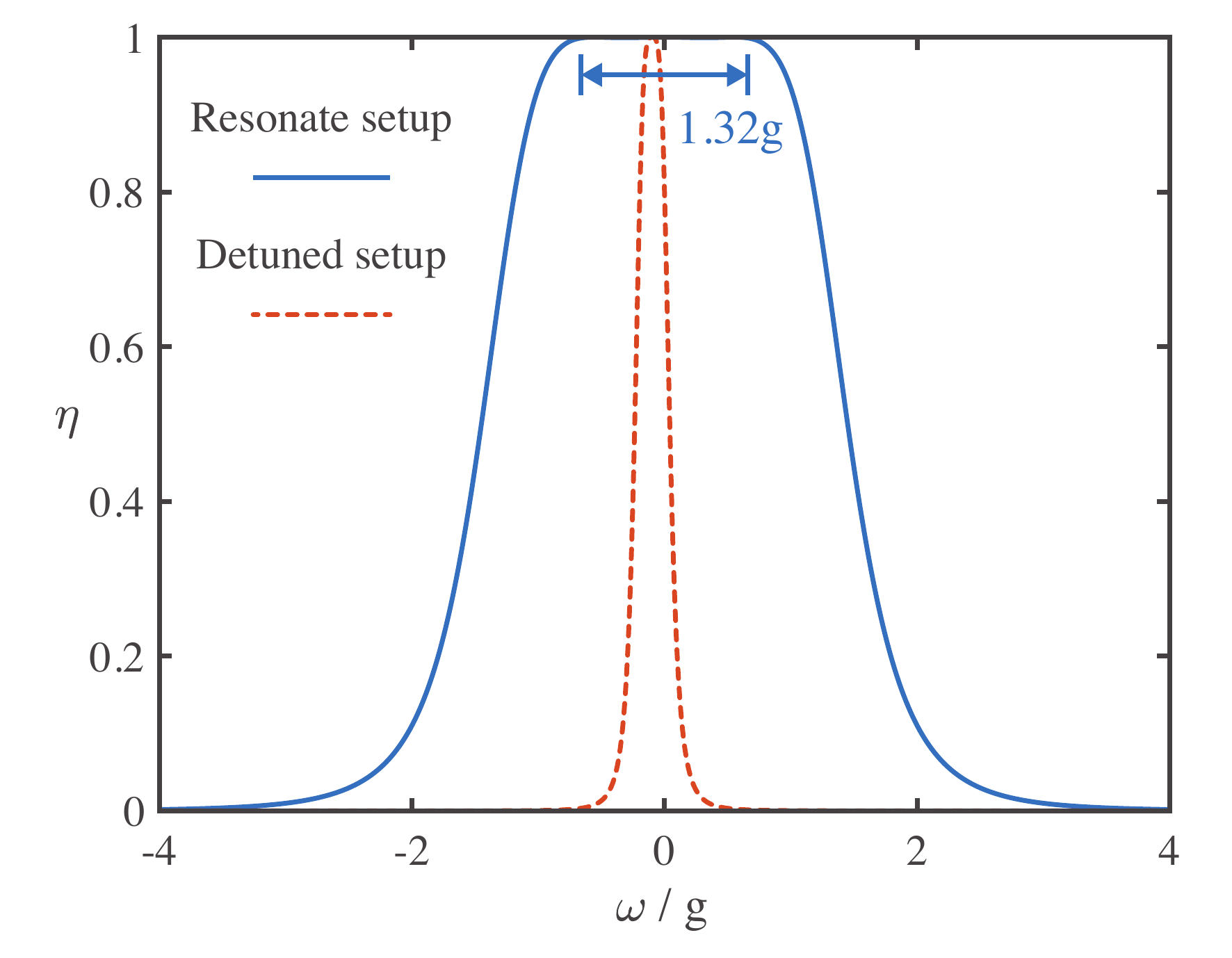}
\caption{(Colour online) The efficiency $\eta$ of the optical field to microwave field converter as a function of the incoming field frequency, with respect to the frequency of the corresponding resonator. When the efficiency reaches one, an incoming optical field can be completely converted into an outgoing microwave field, and vice versa. $g$ is the strength of the effective coupling between the microwave resonator and atoms, which is proportional to $\sqrt{N}$. Here, $N$ is the number of atoms. For the resonant setup with $\kappa = 2.6 g$, the width of the interval with an efficiency higher than $99.9\%$ is $1.32 g$. For the detuned setup, in which we take $\Delta_{\mu} = 10g$ and $\kappa = \frac{2g^2}{\Delta_\mu} = 0.2g$, the bandwidth is much smaller.}
\label{fig:eta}
\end{figure}

In the resonant setup, only the state $\ket{3}$ is off-resonance. When the detuning $\Delta_{o,k}$ is large, the state $\ket{3}$ can be adiabatically eliminated, which leads to the effective coupling between the optical resonator and the transition between $\ket{1}$ and $\ket{2}$. In the resonant setup $\Delta_{\mu,k} = 0$, then Heisenberg equations of motion read
\begin{eqnarray}
\partial_{t} a &=& -i \sum_{k} g_{o,k} \sigma_{13,k}, \label{eq:Heisenberg1} \\
\partial_{t} b &=& -i \sum_{k} g_{\mu,k} \sigma_{12,k}, \label{eq:Heisenberg2} \\
\partial_{t} \sigma_{13,k} &=& -i \Delta_{o,k} \sigma_{13,k} + i g_{o,k} a (\sigma_{33,k}-\sigma_{11,k}) -i \Omega_{k} \sigma_{12,k} \notag \\
&\simeq & -i \Delta_{o,k} \sigma_{13,k} -i g_{o,k} a -i \Omega_{k} \sigma_{12,k}, \label{eq:Heisenberg3} \\
\partial_{t} \sigma_{12,k} &=& i g_{\mu,k} b (\sigma_{22,k}-\sigma_{11,k}) -i \Omega_{k} \sigma_{13,k} \notag \\
&\simeq & -i g_{\mu,k} b -i \Omega_{k} \sigma_{13,k}. \label{eq:Heisenberg4}
\end{eqnarray}
Here, we have assumed that most atoms stay in the state $\ket{1}$, and excitations to states $\ket{2}$ and $\ket{3}$ are few, i.e.~we have replaced $\sigma_{11,k}$ with $1$ and neglected $\sigma_{22,k}$ and $\sigma_{33,k}$. When the detuning $\Delta_{o,k}$ is large, the approximation $\partial_{t}\sigma_{13,k}\simeq 0$ and Eq.~(\ref{eq:Heisenberg3}) yield
\begin{eqnarray}
\sigma_{13,k}=-\frac{g_{o,k}}{\Delta_{o,k}}a-\frac{\Omega_{k}}{\Delta_{o,k}}\sigma_{12,k}.
\label{eq:sigma13}
\end{eqnarray}
Substituting Eq.~(\ref{eq:sigma13}) into Eq.~(\ref{eq:Heisenberg1}) and Eq.~(\ref{eq:Heisenberg4}) gives
\begin{eqnarray}
\partial_{t} a &=& i \sum_{k} \frac{g_{o,k}^{2}}{\Delta_{o,k}} a + i \sum_{k} \frac{g_{o,k}\Omega_{k}}{\Delta_{o,k}} \sigma_{12,k}, \\
\partial_{t} \sigma_{12,k} &=& -i g_{\mu,k} b + i \frac{g_{o,k}\Omega_{k}}{\Delta_{o,k}} a + i \frac{\Omega_{k}^{2}}{\Delta_{o,k}} \sigma_{12,k}.
\end{eqnarray}
These two equations together with Eq.~(\ref{eq:Heisenberg2}) form a new set of Heisenberg equations, which corresponds to the effective Hamiltonian
\begin{eqnarray}
H_{\mathrm{eff}} &=& - \sum_{k} \frac{g_{o,k}^{2}}{\Delta_{o,k}} a^{\dagger}a - \sum_{k} \frac{\Omega_{k}^{2}}{\Delta_{o,k}} \sigma_{22,k} \notag \\
&&- \sum_{k} (\frac{g_{o,k}\Omega_{k}}{\Delta_{o,k}} a^{\dagger} \sigma_{12,k} + h.c.)  \notag\\
&&+ \sum_{k} (g_{\mu,k} b^{\dagger} \sigma_{12,k} + h.c.).
\label{eq:Heff1}
\end{eqnarray}

When excitations are few compared with the number of atoms, we can use boson operators to approximately describe collective excitations, i.e.~$\sum_{k} g_{\mu,k} \sigma_{12,k} \simeq S_{\mu} c$, where $S_{\mu} = \sqrt{\sum_{k} g_{\mu,k}}$ and $c$ is the annihilation operator of bosons. We suppose that the difference between two modes of excitations $\sum_{k} \frac{g_{o,k}\Omega_{k}}{\Delta_{o,k}} \sigma_{12,k}$ and $\sum_{k} g_{\mu,k} \sigma_{12,k}$ are negligible, which is justified when the detuning $\Delta_{o,k}$, coupling strengths $g_{o,k}$ and $g_{\mu,k}$, and the Rabi frequencie $\Omega_{k}$ are approximately uniformly distributed among atoms. Then, $\sum_{k} \frac{g_{o,k}\Omega_{k}}{\Delta_{o,k}} \sigma_{12,k} \simeq S_{o} c$, where $S_{o} = \sqrt{\sum_{k} \frac{g_{o,k}\Omega_{k}}{\Delta_{o,k}}}$. Without loss of generality, we have ignored a potential phase difference between $S_{o}$ and $S_\mu$. Representing atomic excitations using boson operators, the effective Hamiltonian becomes
\begin{eqnarray}
H_{\mathrm{eff}} &=& - S_{o} (a^{\dagger}c + h.c.) + S_{\mu} (b^{\dagger}c + h.c.).
\label{eq:Heff2}
\end{eqnarray}
Here, we have neglected the first two terms in Eq.~(\ref{eq:Heff1}), because they can easily be compensated for by tuning frequencies of the coherent field and resonators. In the next section, we show that the bandwidth of optical field to microwave field converter given by such an effective Hamiltonian is $\sim S_{\mu}$, i.e.~the effective coupling between the microwave resonator and atoms, which can be enhanced by increasing the number of atoms.

\section{Efficiency of the resonant setup}

\begin{figure}[tbp]
\centering
\includegraphics[width=1\linewidth]{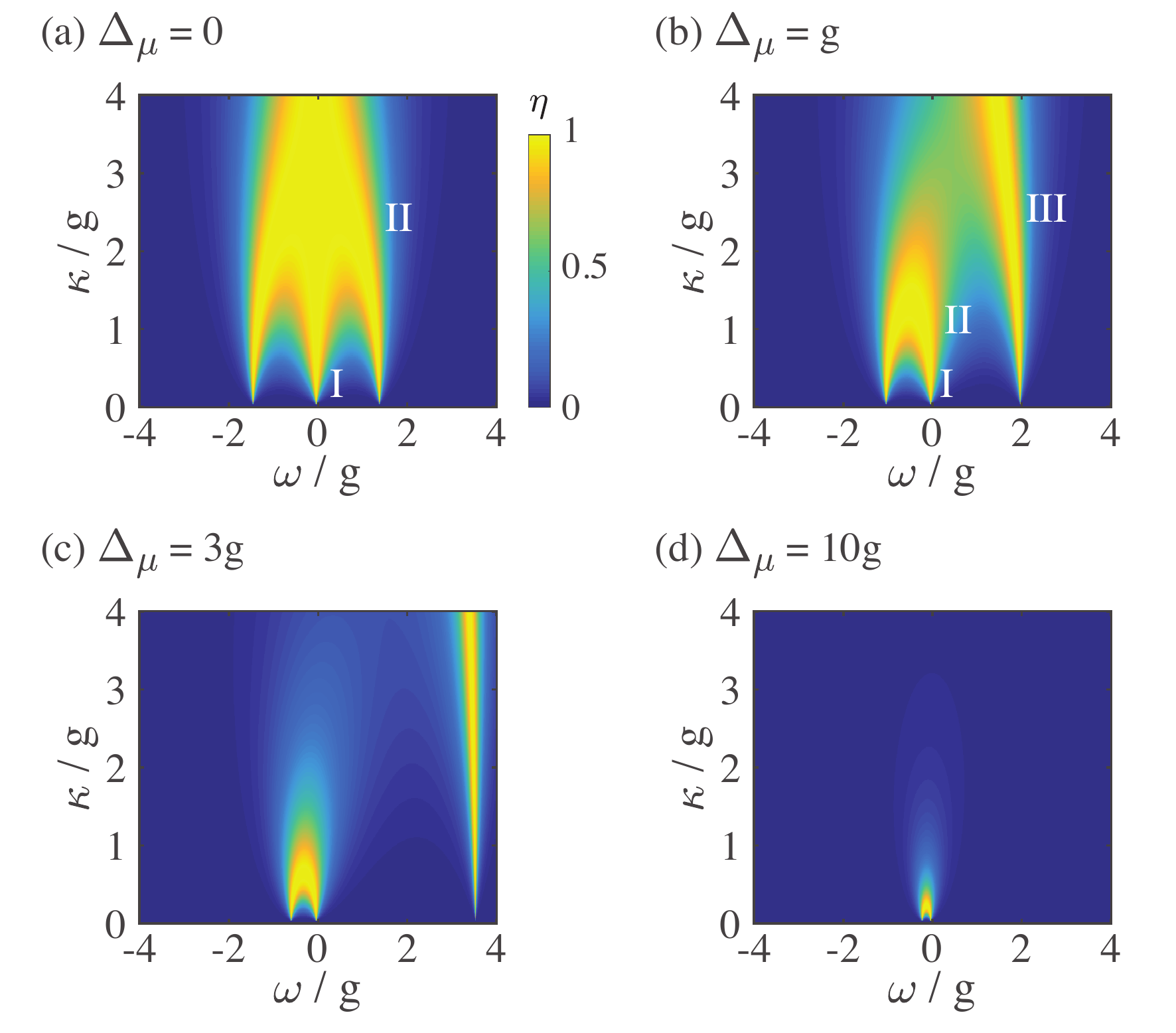}
\caption{(Colour online) The efficiency $\eta$ of the optical field to microwave field converter as a function of the incoming field frequency and the damping rate $\kappa$ for the resonant setup $\Delta_{\mu} = 0$ and detuned setups with $\Delta_{\mu} = g,3g,10g$. The region with high conversion efficiency in the parameter space has three branches when $\kappa$ is low, in which the middle one (marked by I) has the maximum width. When $\kappa$ increases, three (in the resonant setup) or two (in detuned setups) branches merge into one branch (marked by II), which has the maximum width for all values of $\kappa$. In detuned setups, the left branch (marked by III) is disconnected from the other two branches and moves to high frequency when the detuning $\Delta_{\mu}$ increases. When $\kappa$ is high, other two branches disappear, but the left branch remains.}
\label{fig:image}
\end{figure}

In the effective bosonic Hamiltonian of the resonant setup, i.e.~Eq.~(\ref{eq:Heff2}), three boson modes are on-resonance and coupled in the chain structure. In this section, we derive the efficiency of the optical field to microwave field converter. We find that the efficiency can reach unity, i.e.~optical photons resonating with the optical resonator can be completely converted into microwave photons and the other way around. The efficiency gently decreases if the photon frequency is off-resonance, and the bandwidth is $\sim S_{\mu}$.

We consider an $n$-mode problem and start with the quantum Langevin equation for the $i$-th mode, which is~\cite{collett:84}
\begin{eqnarray}
\dot{a}_{i} = -i \sum_{j} A_{ij} a_{j} - \frac{\kappa_{i}}{2} a_{i} - \Gamma_{i}, \label{eq:dota}
\end{eqnarray}
where $\dot{a}_{i} = \frac{da}{dt}$, $A_{i,j}$ is the coupling strength between the $i$-th mode and the $j$-th mode, $\kappa_{i}$ is the damping rate of the $i$-th mode, and $\Gamma_{i}=\sqrt{\kappa_{i}}a_{i}^{(\mathrm{in})}$ is the inhomogeneity with $a_{i}^{\mathrm{(in)}}$ the incoming field of the $i$-th mode. If we define the matrices as
\[ \bar{a} = \left( \begin{array}{ccc}
a_{1} \\
a_{2} \\
\vdots \\
a_{n} \end{array} \right),
K= \left( \begin{array}{cccc}
\kappa_{1} & 0 & 0 & 0 \\
0 & \kappa_{2} & 0 & 0  \\
0 & 0 & \ddots & 0  \\
0 & 0 & 0 & \kappa_{n} \end{array} \right), \]
the quantum Langevin equation of $n$ modes takes the form
\begin{eqnarray}
\dot{\bar{a}} = -i A\bar{a} - \frac{K}{2} \bar{a} - \sqrt{K} \bar{a}^{(\mathrm{in})}.
\label{eq:aain}
\end{eqnarray} 
Under the time reversal transformation, $t\rightarrow -t$, the incoming field in Eq.~(\ref{eq:aain}) is replaced by the outgoing field $\bar{a}^{\mathrm{(out)}}$, and the sign in the systematic part changes~\cite{collett:84}. Thus Eq.~(\ref{eq:aain}) becomes 
\begin{eqnarray}
- \dot{\bar{a}} = i A\bar{a} - \frac{K}{2} \bar{a} - \sqrt{K} \bar{a}^{(\mathrm{out})}.
\label{eq:aaout}
\end{eqnarray} 
Equations~(\ref{eq:aain})~and~(\ref{eq:aaout}) together give
\begin{eqnarray}
\bar{a}^{(\mathrm{out})} = - \sqrt{K} \bar{a} - \bar{a}^{(\mathrm{in})}.
\label{eq:aout}
\end{eqnarray}
We now express Eq.~(\ref{eq:aain}) in the form of frequency components, yielding
\begin{eqnarray}
-i \omega \bar{a}(\omega) = -i A \bar{a}(\omega) - \frac{K}{2} \bar{a}(\omega) - \sqrt{K} \bar{a}^{(\mathrm{in})}(\omega),
\end{eqnarray}
and thus
\begin{eqnarray}
\bar{a}(\omega) = -2 (K-2i\omega+2iA)^{-1} \sqrt{K} \bar{a}^{(\mathrm{in})}(\omega).
\label{eq:a}
\end{eqnarray} 
Substituting Eq.~(\ref{eq:a}) into Eq.~(\ref{eq:aout}), the output field is then given by
\begin{eqnarray}
\bar{a}^{(\mathrm{out})} = \left[ 2 \sqrt{K} (K-2i\omega+2iA)^{-1} \sqrt{K} - \openone \right] \bar{a}^{(\mathrm{in})},
\label{eq:finalaout}
\end{eqnarray}
where $\openone$ is the identity operator. This equation describes the relation between input and output fields.

Given our effective bosonic Hamiltonian of the resonant setup, i.e.~Eq.~(\ref{eq:Heff2}), and taken $\bar{a} = ( a~c~b)^{\rm T}$, the corresponding matrices are
\[ A= \left( \begin{array}{ccc}
0 & S_{o} & 0  \\
S_{o} & 0 & S_{\mu}  \\
0 & S_{\mu} & 0  \end{array} \right) ,
K= \left( \begin{array}{ccc}
\kappa_{o} & 0 & 0  \\
0 & 0 & 0  \\
0 & 0 & \kappa_{\mu} \end{array} \right). \]
Here, $\kappa_{o}$ is the damping rate of the optical resonator, and $\kappa_{\mu}$ is the damping rate of the microwave resonator. From now on we focus on the case that $S_{o} = S_{\mu} = g$ and $\kappa_{0} = \kappa_{\mu} = \kappa$.

We denote $a_{\pm} = \frac{1}{\sqrt{2}} (a \pm b)$, then quantum Langevin equations become
\begin{eqnarray}
\dot{a}_{-}& = & - \frac{\kappa}{2} a_{-} - \sqrt{\kappa} a_{-}^{(\mathrm{in})}, \notag \\
\dot{c} &=& - \sqrt{2} ig a_{+}, \notag \\
\dot{a}_{+} &=& - \sqrt{2} ig c - \frac{\kappa}{2} a_{+} - \sqrt{\kappa} a_{+}^{(\mathrm{in})}.
\end{eqnarray}
According to these equations, the mode $a_{-}$ is decoupled from other two modes. Therefore, the incoming field of the mode $a_{-}$ is directly reflected, but the incoming field of the mode $a_{+}$ is reflected via the coupling to the mode $c$. Because of the existence of the mode $c$, the outgoing fields of $a_{-}$ and $a_{+}$ have a phase difference. When the phase difference is $\pi$, photons are converted from $a$ into $b$ or the other way around. From Eq.~(\ref{eq:finalaout}), we have
\begin{eqnarray}
a_{-}^{(\mathrm{out})} = \frac{\kappa+i2\omega}{\kappa-i2\omega} a_{-}^{(\mathrm{in})}
\label{eq:a-}
\end{eqnarray}
and
\begin{eqnarray}
a_{+}^{(\mathrm{out})} = \frac{\kappa+2i\omega-i4g^{2}/\omega}{\kappa-2i\omega+i4g^{2}/\omega} a_{+}^{(\mathrm{in})}.
\label{eq:a+}
\end{eqnarray}
Equations~(\ref{eq:a-})~and~(\ref{eq:a+}) together give
\begin{eqnarray}
a^{(\mathrm{out})} &=& \frac{1}{2} \left[ \frac{\kappa+2i\omega}{\kappa-2i\omega} (a^{(\mathrm{in})}-b^{(\mathrm{in})}) \right. \notag \\ 
&&+ \left. \frac{(\kappa+2i\omega)\omega-i4g^{2}}{(\kappa-2i\omega)\omega+i4g^{2}} (a^{(\mathrm{in})}+b^{(\mathrm{in})}) \right], \\
b^{(\mathrm{out})} &=& \frac{1}{2} \left[ \frac{\kappa+2i\omega}{\kappa-2i\omega} (b^{(\mathrm{in})}-a^{(\mathrm{in})}) \right. \notag \\ 
&&+ \left. \frac{(\kappa+2i\omega)\omega-i4g^{2}}{(\kappa-2i\omega)\omega+i4g^{2}} (b^{(\mathrm{in})}+a^{(\mathrm{in})}) \right].
\end{eqnarray}
Here, $a^{(\mathrm{in})}$/$a^{(\mathrm{out})}$ and $b^{(\mathrm{in})}$/$b^{(\mathrm{out})}$ are input/output fields of the optical resonator and microwave resonator, respectively. The optical field to microwave field conversion efficiency is defined as
\begin{eqnarray}
\eta &=& \vert a^{(\mathrm{out})}/b^{(\mathrm{in})} \vert ^{2} = \vert b^{(\mathrm{out})}/a^{(\mathrm{in})} \vert ^{2} \notag \\
&& = \frac{1}{4} \left\vert \frac{\kappa+2i\omega}{\kappa-2i\omega} - \frac{(\kappa+2i\omega)\omega-i4g^{2}}{(\kappa-2i\omega)\omega+i4g^{2}} \right\vert^2.
\end{eqnarray}
When $\kappa = 2g$, the efficiency $\eta = 1$ at frequencies $\omega = 0,\pm g$.

\section{Bandwidths of setups}

\begin{figure}[tbp]
\centering
\includegraphics[width=1\linewidth]{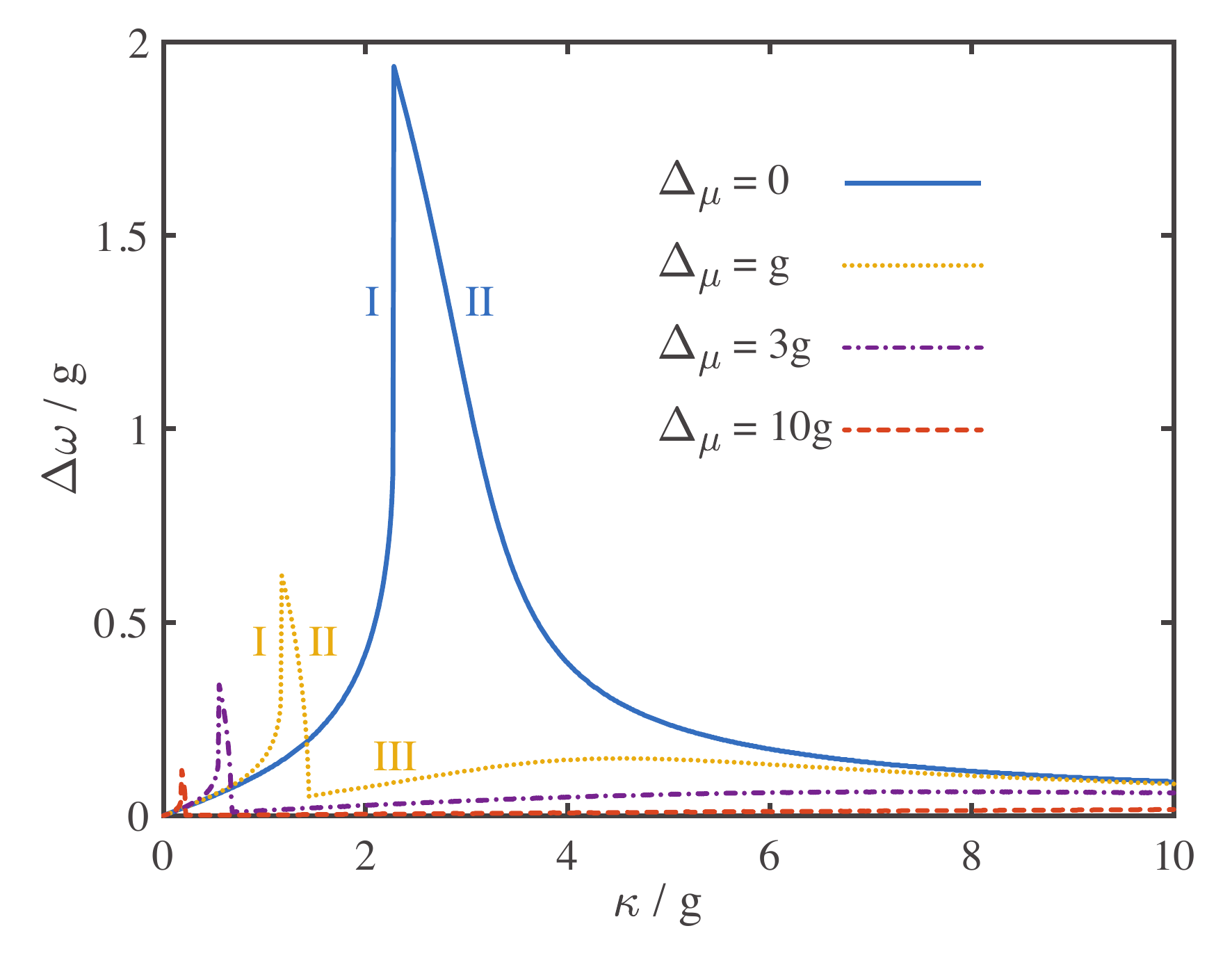}
\caption{(Colour online) The maximum width $\Delta\omega$ of an interval in the frequency space, in which the conversion efficiency $\eta$ is higher than $99\%$. The branch with the maximum width (see Fig.~\ref{fig:image}) is marked for the resonant setup $\Delta_{\mu} = 0$ and the detuned setup $\Delta_{\mu} = g$. Other two detuned setups $\Delta_{\mu} = 3g,10g$ are similar.}
\label{fig:width}
\end{figure}

For the resonant setup, the conversion efficiency equals to $1$ at $\omega=0$, and the typical bandwidth is $\sim g$. We remark that $g = S_{\mu} = \sqrt{\sum_{k} g_{\mu,k}}$ is the effective coupling between the microwave resonator and atoms. In Fig.~\ref{fig:eta}, the efficiency with $\kappa = 2.6 g$ is plotted, and we can find a broad flat around the frequency $\omega = 0$, which implies a big bandwidth. In such a setup, the efficiency is greater than $99.9\%$ in a frequency interval with the width $1.66 g$.

To compare with the detuned setup, we consider the effective Hamiltonian
\begin{eqnarray}
H_{\mathrm{eff}}' &=& \Delta_\mu c^{\dagger}c -S_{o}(a^{\dagger}c+h.c.) +S_{\mu} (b^{\dagger}c+h.c.),
\end{eqnarray}
where $\Delta_\mu$ denotes the detuning of the transition between $\ket{2}$ and $\ket{3}$, and $S_{o} = S_{\mu} = g$. When $\Delta_\mu \gg g$, the adiabatic elimination of the mode $c$ leads to the effective Hamiltonian $H_{a,b}$ with $S = \frac{g^2}{\Delta_\mu}$. The conversion efficiency can research $1$ when $\kappa_o = \kappa_\mu = 2S$. In Fig.~\ref{fig:eta}, the efficiency of the detuned setup is plotted taking $\Delta_\mu = 10 g$. We can find that the bandwidth is much smaller.

The resonant setup is optimal compared with all detuned setups concerning the bandwidth. In Fig.~\ref{fig:image}, we plot the conversion efficiency as a function of the frequency $\omega$ and the damping rate $\kappa_o = \kappa_\mu = \kappa$ for $\Delta_\mu = 0, g, 3g, 10g$. We can find that in the parameter space the area of high-efficiency conversion is smaller when the detuning is larger. For the resonant setup, there are three branches of the high-efficiency area at low damping rate, which merge into one branch when the damping rate increases. When the detuning is switched on, the left branch becomes disconnected from the other two branches, which moves toward the high-frequency direction when the detuning increases. When the detuning $\Delta_\mu$ is much larger than $g$, only a small high-efficiency area around the zero frequency left. In Fig.~\ref{fig:width}, the bandwidth, i.e.~the maximum width of a frequency interval with an efficiency larger than $99\%$, is plotted. The maximum bandwidth decreases with the detuning, and for the resonant setup, the maximum bandwidth can reach $\sim 2g$.

\section{Conclusions}

In this work, we propose that the conversion bandwidth can be largely increased if we consider the setup with a large optical transition detuning $\vert \Delta_{o} \vert \gg \vert \sqrt{N}g_{o}\vert$ and a vanising microwave transition detuning $\Delta_{\mu}=0$. Working with a large $\Delta_{o}$ enables the adiabatic elimination of the optical excited state. Then the system becomes an effective system with the optical and microwave modes both coupled to the atomic excitation in the metastable state. The effective coupling strengths are $\sqrt{N}\abs{g_{o}\Omega/\Delta_{o}}$ and $\sqrt{N}\abs{g_{\mu}}$, respectively. They are much larger than the effective coupling $N\abs{\Omega g_{\mu} g_{o}^*/(\Delta_{\mu} \Delta_{o})}$ obtained in Ref. \cite{williamson:14}, which implies a much higher bandwidth. Although two resonators are not directly coupled in our effective model, the conversion efficiency can still reach $100\%$, and the efficiency is almost a constant within a large frequency interval with the typical width $\sqrt{N}\abs{g_{\mu}}$. These results show that a very high conversion bandwidth and short conversion time can be achieved using the resonant setup.

\section*{acknowledgements}

The author acknowledges the optical and microwave fields conversion project suggested by Simon Benjamin. The author also acknowledges the discussions with Joshua Nunn and Peter Leek. I am also grateful to Ying Li for his help in preparing the manuscript. This work is supported by the National Key R$\&$D Program of China grant 2016YFA0301200. It is also supported by Science Challenge Project (under Grant No. TZ2018003), the National Natural Science Foundation of China (under Grant No. U1530401), and China Postdoctoral Science Foundation.

\end{document}